\documentclass[12pt,preprint]{aastex}

\newcommand\simlt{\lower.5ex\hbox{$\; \buildrel < \over \sim \;$}}

\slugcomment{ApJ, to be submitted}

\shorttitle{ICS close to the upper cut-off}
\shortauthors{E. Lefa et al.}

\begin{document}

\title{On the spectral shape of radiation due to Inverse Compton Scattering close to the maximum cut-off}

\author{E. Lefa\altaffilmark{1,*}}
\affil{$^1$ Max-Planck-Institut f\"ur Kernphysik, P.O. Box 103980, 69029 Heidelberg, Germany}
\email{eva.lefa@mpi-hd.mpg.de}
\author{S. R. Kelner\altaffilmark{1}}
\affil{$^1$ Max-Planck-Institut f\"ur Kernphysik, P.O. Box 103980, 69029 Heidelberg, Germany}
\and
\author{F. A. Aharonian\altaffilmark{1,2}}
\affil{$^1$ Dublin Institute for Advanced Studies, 31 Fitzwilliam Place, Dublin 2, Ireland}
\affil{$^2$ Max-Planck-Institut f\"ur Kernphysik, P.O. Box 103980, 69029 Heidelberg, Germany}

\let\oldthefootnote\thefootnote
\renewcommand{\thefootnote}{\fnsymbol{footnote}}
\footnotetext[1]{Fellow of the International Max Planck Research
    School for Astronomy and Cosmic Physics at
                   the University of Heidelberg (IMPRS-HD)}
\let\thefootnote\oldthefootnote

\begin{abstract}
The spectral shape of radiation due to Inverse Compton Scattering is analyzed,
in the Thomson and the Klein-Nishina regime, for electron distributions with exponential cut-off.
We derive analytical, asymptotic expressions
for the spectrum close to the maximum cut-off region. We consider monoenergetic,
Planckian and Synchrotron photons as target photon fields.
These approximations provide a direct link between the distribution of parent electrons
and the up-scattered spectrum at the cut-off region.
\end{abstract}

\keywords{Radiation mechanisms: non-thermal -- Scattering -- Gamma rays: galaxies  -- }

\newpage

\section{Introduction}
The interaction of relativistic electrons with low energy radiation through Inverse Compton Scattering
(ICS) provides one of the principal $\gamma$-ray production processes in astrophysics.
In a variety of astrophysical environments, from very compact objects like pulsars and
Active Galactic Nuclei (AGN) to extended sources like supernova remnants and clusters
of galaxies, low energy photons are effectively boosted to high energies through this mechanism.

The basic features of the ICS have been analyzed by \cite{jones68}, \cite{BG70}.
The case of anisotropic electrons and/or photons has been studied by \cite{AA81c},
\cite{narginen93}, \cite{brunetti00} and \cite{sazonov00}. The
impact  of the Klein-Nishina effect on the formation of the energy distribution of
electrons  was  first  realized by \cite{blumen71}. Its importance in astrophysics has been
discussed in the literature in the context of different non thermal phenomena, in particular by \cite{Ahar85}, \cite{zdziarski89}, \cite{dermer02}, \cite{sikora05}, \cite{mitia}, \cite{kusunose05}, \cite{stawarz06} and \cite{stawarz2010}.

Generally, the energy spectrum as well as the effects related to ICS are numerically
calculated using the exact expression for the Compton cross section. On the other hand,
compact, analytical approximations can serve as useful tools for a deeper understanding
of the properties of Compton radiation and the implications of the complex numerical calculations.
In particular, inferring the energy distribution of the parent particles from the observed
spectrum is a much more efficient procedure when analytical approximations are available.
For example, a power-law distribution of electrons normally results in power-law photon spectra.
If the observed photon index is $\Gamma$ (in a $dN_{\gamma}/dE_{\gamma}\propto E_{\gamma}^{-\Gamma}$
representation), then we can obtain the power-law index of the electron distribution
$dN_e/dE_e\propto E^{-\Gamma_{e}}$ from the relation $\Gamma_e=2\Gamma-1$ for the Thomson regime
and $\Gamma_e\approx \Gamma-1$ if the scattering occurs in the Klein-Nishina regime.

This however only applies to the energy interval far from the cut-off, the "main" part of the electron distribution.
At the highest (and lowest) energies, there should
be a break/cut-off in the electron distribution and in fact, the corresponding break at the radiation
spectrum contains a lot of interesting information on the parent electrons.
In particular, the peaks in the Spectral Energy Distribution (SED)
appear at this energy range in the majority of cases, indicating that the source luminosity
is mostly radiated at the maximum cut-off. Moreover, as the main, power-law part of the distribution,
the shape of the cut-off carries as well important implication for the acceleration of the particles
and in general the mechanisms acting in the source. Although the shape of the spectrum close to the highest
energy cut-off is critical, this topic has not yet been adequately addressed. In this paper
we examine the shape of the Compton spectrum close to the maximum cut-off and
we derive convenient analytical formulas that allow to approximate the radiated flux in this
specific energy range.

In general, the shape of the electron distribution around the cut-off can be expressed as
an exponential, $exp[-(E_e/E_{c})^\beta]$. This allows us to describe a quite broad
range of distributions, even very sharp, abrupt, step-function like cut-offs for
$\beta\gg 1$. Apart from the convenience of such a mathematical description, exponential
cut-offs naturally arise in theoretical considerations. For example, in diffusive shock
acceleration, power-law particle distributions with exponential cut-off are formed when (synchrotron)
energy losses are taken into account \cite{webb84} and
the cut-off index is very important for investigating the acceleration mechanism.
Analytic solutions show that in the case of Bohm diffusion a simple exponential cut-off
$\exp{[-E_e/E_c]}$ arises, whereas the index approaches $\beta=2$ if
$\dot{E_e}\propto E^{2}_{e}$ type  energy losses are taken into account,
e.g. synchrotron or Thomson losses (see \citealt{zirak07}).

In stochastic acceleration scenarios, where pile-up particle distributions are formed when
acceleration is balanced by synchrotron type losses, the shape of the electron cut-off is
directly related to the assumed turbulent wave spectrum (\citealt{Schlickeiser85},
\citealt{aharonian86}), e.g. $\beta=5/3$ for Kolmogorov, $\beta=3/2$ for Kraicman like or $\beta=2$
for the hard sphere approximation. Of course, if more complicated energy losses dominate,
like in the case of Klein-Nishina losses in radiation-dominated environments, more complex shapes for the
electron distribution cut-off may be expected in both stochastic and diffusive
shock acceleration scenarios (e.g. \citealt{lukas2008}, \citealt{vannoni2009})

Nevertheless, it seems reasonable to consider particle distributions that exhibit exponential
cut-off in a general form, for investigating and modeling the radiated spectra.
In \cite{frit89} and \cite{zirak07}, the shape of the synchrotron spectrum close
to the high energy cut-off has been discussed. They found
that when the electron distribution possesses an exponential cut-off of
index $\beta$, $\exp\left[{-(E_{e}/E_c)^{\beta}}\right]$, then the radiated synchrotron
spectrum exhibits a smoother cut-off, of index $\beta/(\beta+2)$. Apart from the practical
importance, this analytic result demonstrates that a $\delta$-function approximation for the
synchrotron radiation emissivity does not give the correct result.

Here we examine the corresponding Compton spectrum, in the Thomson and Klein-Nishina regimes,
considering different target photon fields so that both synchrotron self-Compton (SSC)
and external Compton (EC) scattering can be addressed. We derive analytically the asymptotic
behavior of the up-scattered photon distribution close to the cut-off region.
We consider a general electron energy distribution of the form
\begin{equation}\label{electrons}
\frac{dN_{e}}{dE_e}=F_{e}(E_{e})=A E^{\alpha}_{e}e^{-\left(\frac{E_e}{E_c}\right)^\beta},
\end{equation}
where $E_e=\gamma mc^2$ is the electron energy and $E_{c}=\gamma_c mc^2$ is the cut-off energy.
This presentation allows us to consider either a power-law distribution (for $\alpha=-|\alpha|$)
with exponential cut-off or a relativistic Maxwell-like distribution (for $\alpha=2$), that may
be formed in stochastic acceleration scenarios. We consider monochromatic,
Planckian and synchrotron photons as target photon fields. The resulting IC spectral shape
is discussed for the Thomson and the Klein-Nishina regime and we show that is not always
identical to the synchrotron spectrum, as is often silently assumed. Finally, we discuss basic
features and physical properties of the radiated spectrum.

\section{Compton spectrum for monochromatic photons}
In this section we calculate the asymptotic behavior that the Compton spectrum exhibits
close to the maximum cut-off, when monochromatic photons (with isotropic angular distribution)
are up-scattered. Let as consider the general function of eq. ($\ref{electrons}$),
that describes the differential number of electrons. Electrons are considered to be
isotropically and homogeneously distributed in space. Then the spectrum of photons
generated per unit time due to ICS is (see e.g. \citealt{BG70})
\begin{equation}\label{icm2}
d\dot{N}_{\gamma}/dE_{\gamma}=\int^{\infty}_{0}\int^{\infty}_{E_{e min}}W(E_{e},\epsilon_{\gamma},E_{\gamma})F_{e}(E_{e})n_{ph}(\epsilon_{\gamma})dE_{e}d\epsilon_{\gamma},
\end{equation}
where
\begin{equation}\label{icm3}
E_{e\min}=\frac12\,E_\gamma\left(1+\sqrt{1+\frac{m^2 c^4}{\epsilon_\gamma E_\gamma}}\right),
\end{equation}
\begin{equation}\label{icm4}
W(E_e,\epsilon_\gamma,E_\gamma)=
\frac{8\pi r_e^2 c}{E_e\,\eta}\left[ 2q\,\ln q+(1-q)\left(1+2q+
\frac{\eta^2q^2}{2\,(1+\eta q)}\right)\right],
\end{equation}
and
\begin{equation}\label{icm5}
\eta=\frac{4\,\epsilon_\gamma E_e}{m^2 c^4},\quad q=\frac{E_\gamma}{\eta\,(E_e-E_\gamma)}\;.
\end{equation}
Here the function $W(E_e,\epsilon_{\gamma},E_{\gamma})$
in eq. (\ref{icm4}) describes the total scattering probability, taking into account Klein-Nishina effects.
The parameter $\eta$ in eq. (\ref{icm5}) defines the domain of the scattering.
For $\eta<<1$ the Thomson regime applies whereas for $\eta\gg 1$ we are in the Klein Nishina regime.
In the case of monochromatic photons, the number density  is $n_{ph}(\epsilon_{\gamma})=n_0\delta(\epsilon_{\gamma}-\varepsilon_{0})$ and we will set $n_0$ equal to 1 for this case.
From now on we set $mc^2=1$ throughout the calculations for simplicity, apart from the formulas at which
the results are demonstrated. The case of $\alpha<0$ will be referred to as power-law distribution,
whereas $\alpha=2$ will be referred to as Maxwellian distribution.

\subsection{Thomson Regime}

In the limiting case of $4\varepsilon_{0} E_{\rm c}\ll 1$,
when all the scatterings occur in the Thomson regime, the photons
take a small fraction of the electron energy. Thus, from the previous
relation, it follows that $E_\gamma\ll 1/(4\varepsilon_{0})$ and the lower
limit of the integration becomes
\begin{equation}\label{icm6}
E_e\ge E_{e\min} \approx \sqrt{E_\gamma/(4\varepsilon_{0})}\gg
E_\gamma\,.
\end{equation}
Therefore, in this case $\eta q=E_\gamma/E_e\ll 1$ and eq. (\ref{icm4}) can be written as
\begin{equation}\label{icm7}
W(E_e,\varepsilon_{0},E_\gamma)=
\frac{8\pi r_e^2 c}{E_{\gamma\max}}\left[ 2q\,\ln q+(1-q)\left(1+2q
\right)\right],
\end{equation}
and in this approximation $q=E_\gamma/E_{\gamma\max}$, where
$E_{\gamma\max}=4\varepsilon_{0} E_e^2$. Using the electron distribution of eq. (\ref{electrons})
and changing the integration variable from $E_{e}$ to $q$, the integral of eq. (\ref{icm2}) for the
Compton spectrum becomes
\begin{equation}
d\dot{N}_{\gamma}/dE_{\gamma}=\int^{1}_{0}\frac{2\pi r_e^2 c}{2^{\alpha}}\varepsilon^{-\frac{\alpha+1}{2}}_{0}E^{\frac{\alpha-1}{2}}_{\gamma}e^{-\frac{\xi}{q^{\beta/2}}}q^{-\frac{\alpha+1}{2}}\left[ 2q\,\ln q+(1-q)\left(1+2q
\right)\right]dq,
\end{equation}
where
\begin{equation}\label{th4}
\xi=\left(\frac{E_\gamma}
{4\varepsilon_0 E_{\rm c}^2}\right)^{\!\beta/2}\,.
\end{equation}
We are interested in the behavior of the spectrum near the
exponential cut-off, i.e. $E_{\gamma}>>4\varepsilon_{0}E^{2}_{c}$
or $\xi>>1$. Then the integrant is dominated by values of $q$ very
close to unity and in order to perform the above integration it is
convenient to change again variables to $\tau=q^{-\beta/2}$ so that
\begin{equation}\label{icm8}
d\dot{N}_{\gamma}/dE_{\gamma}=\int^{\infty}_{1}\frac{2\pi r_e^2 c}{2^{\alpha}}\;\varepsilon^{-\frac{\alpha+1}{2}}_{0}E^{\frac{\alpha-1}{2}}_{\gamma}f(\tau)e^{-\xi\tau}d\tau,
\end{equation}
where
\begin{equation}
f(\tau)=2\frac{4\tau^{\frac{\alpha-3}{\beta}}\ln{\tau}-\beta\tau^{\frac{\alpha-1}{\beta}}-\beta\tau^{\frac{\alpha-3}{\beta}}+2\beta\tau^{\frac{\alpha-5}{\beta}}}{\beta^2\tau}.
\end{equation}

\begin{figure}[!h]
\epsscale{0.3pt}
\includegraphics[width=180pt,height=290pt,angle=-90]{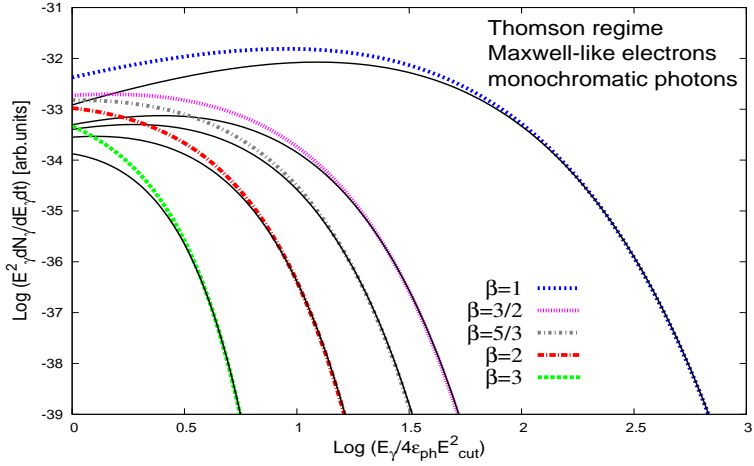}
\caption{Asymptotics in the Thomson regime for monochromatic photons. The $\eta_{c}=4\varepsilon_0 E_c$ parameter that defines the domain of scattering for the cut-off electrons is $\eta_{c}=0.0004$. $E_c=10^2$ and $\varepsilon_0=10^{-6}$ both in $mc^2$ units. For this figure relativistic Maxwell-like electron distributions are used with different shapes of exponential cut-off, $\beta$ parameter. The exponential cut-off of the up-scattered photon spectrum posses an index of $\beta/2$. As in all figures, black solid lines show the numerical spectrum, whereas colored, dashed lines correspond to the analytic approximation.}\label{f1T}
\vskip\belowcaptionskip
\end{figure}

\begin{figure}[!h]
\epsscale{0.3pt}
\includegraphics[width=190pt,height=290pt,angle=-90]{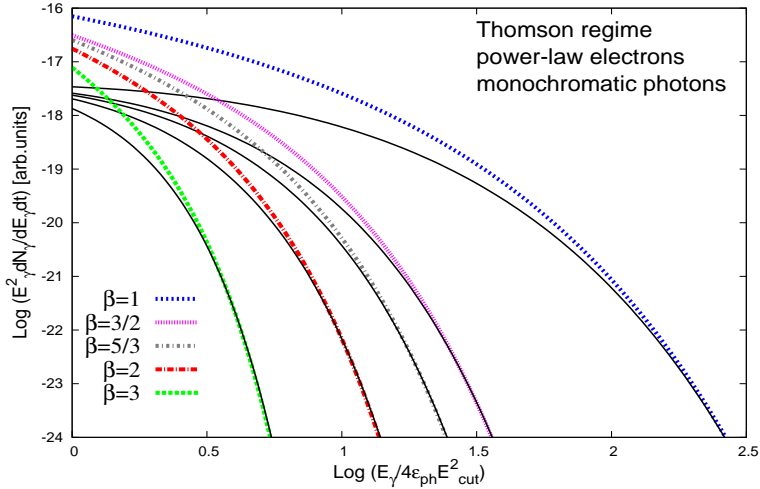}
\caption{Same as figure \ref{f1T} but for power-law electrons. The asymptotics approach the numerical solution only for $E_{\gamma}\gg 4\varepsilon_0 E^{2}_{cut}$. For smaller values of $E_{\gamma}$ the numerical spectrum is smoother than the approximated one.
}\label{f2T}
\vskip\belowcaptionskip
\end{figure}

The function $f(\tau)$ is also dominated by values
of $\tau$ close to unity. Thus, we expand $f(\tau)$
in series around $\tau=1$. By keeping terms up to first order,
the resulting spectrum reads
\begin{equation}\label{th3}
 \left.\frac{d\dot N_\gamma}{dE_\gamma}\right|^{\;T}=\frac{8\pi r_e^2c A(mc^2)^{\alpha+1}}{2^{\alpha}}\,
\frac{\varepsilon^{-\frac{\alpha+1}{2}}_{0} E^{\frac{\alpha-1}{2}}_{\gamma}}{\beta^2\xi^2}
e^{-\xi}\,, \quad {\rm for}  \quad
E_\gamma\gg\frac{\epsilon_\gamma E_{\rm c}^2}{(mc^2)^2}\,.
\end{equation}
The above expression gives the asymptotic behavior of
the Thomson spectrum with exponential accuracy at the cut-off region.
When monochromatic photons are up-scattered by electrons with index $\beta$,
the radiated flux exhibits a smoother cut-off, of index $\beta/2$, i.e.
\begin{equation}\label{beta2}
\left.\frac{d\dot N_\gamma}{dE_\gamma}\right|^{\;T}\propto \exp\left[-\left(\frac{E_{\gamma}(m c^2)^2}{4\varepsilon_{0} E^{2}_{c}}\right)^{\beta/2}\right].
\end{equation}
As expected, the cut-off in the photon spectrum, $4\varepsilon_{0} E^{2}_{c}/(m c^2)^2$,
corresponds to the maximum photon energy that an electron of energy $E_c$ can radiate in the Thomson regime.
We also note that in this case, an abrupt cut-off ($\beta\rightarrow\infty$)
of the electron energy distribution would correspond to an abrupt cut-off of the photon spectrum.
In figures \ref{f1T} and \ref{f2T} the analytic formula of eq. (\ref{th3}) and the full,
numerical spectrum are plotted, for a Maxwellian and a power-law distribution respectively.
Asymptotics are better for $\alpha=2$. For the power-law distribution the numerical and analytical solution
converge for very large values of the parameter $\xi$. In both cases, very close to the cut-off energy,
the numerical spectrum is smoother than the approximated one.

Finally, for a "pure" power-law distribution without
exponential cut-off ($\alpha$ negative and $\beta=0$) one
can integrate eq. (\ref{icm8}) directly to obtain the result
\begin{equation}\label{icm9}
\left.\frac{d\dot N_\gamma}{dE_\gamma}\right|^{\;T}=\frac{2\pi r_e^2 cA(mc^2)^3}{2^{\alpha}}\varepsilon^{-\frac{\alpha+1}{2}}_{0}E^{\frac{\alpha-1}{2}}_{\gamma}\frac{4(\alpha^2-4\alpha+11)}{(\alpha-3)^2(\alpha^2-6\alpha+5)}, \quad {\rm if}  \quad \alpha<3,
\end{equation}
that demonstrates that in the Thomson regime the radiated spectrum
follows a power-law of the form $d\dot N_\gamma/dE_\gamma\propto E^{\frac{\alpha-1}{2}}_{\gamma}$.

\subsection{Klein-Nishina Regime}

\begin{figure}[!h]
\epsscale{0.3pt}
\includegraphics[width=190pt,height=290pt,angle=-90]{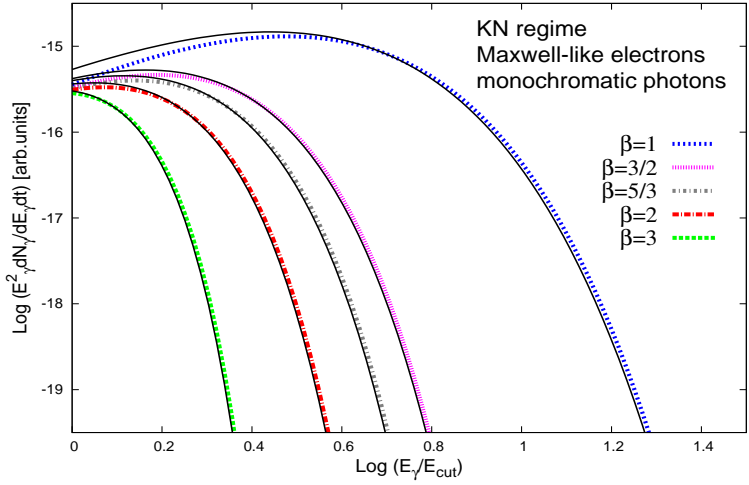}
\caption{Asymptotics in the Klein-Nishina regime for monochromatic photons and Maxwell-like electrons. The $\eta_c$ parameter that defines the domain of the scattering is $\eta_c=400$. $E_c=10^6$ and $\varepsilon_0=10^{-4}$ both in $mc^2$ units. The photon spectrum exhibits a cut-off index $\beta$, same as the electron distribution. The analytical approximation is in very good agreement with the numerical spectrum just after the peak of the SED.}\label{f1KN}
\end{figure}

\begin{figure}[!h]
\epsscale{0.3pt}
\includegraphics[width=190pt,height=290pt,angle=-90]{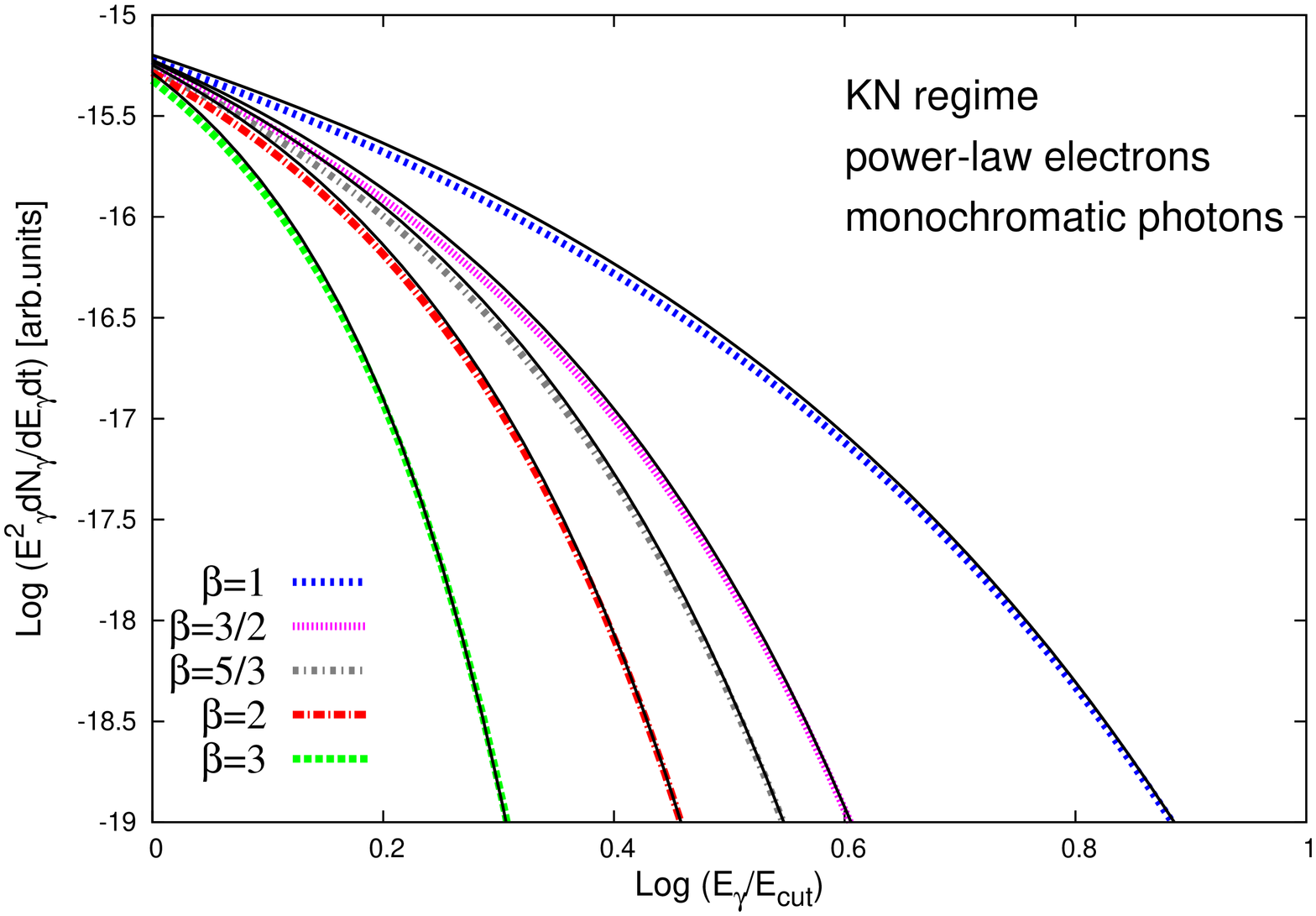}
\caption{Same as figure \ref{f1KN} but for a power-law electron distribution.}\label{f2KN}
\end{figure}

\noindent When $\eta=4\varepsilon_{0}E_{e}\gg 1$, photons take
almost all the energy of the electrons in one scattering. Then, we
may define the parameter $\zeta\equiv 4\varepsilon_{0}E_{\gamma}$
and the low limit of the integration in eq. (\ref{icm3}) becomes
$E_{emin}\approx E_{\gamma}$. Let us change variables to
$x\equiv E_{\gamma}/E_e$. As $\zeta\gg 1$, we keep only the leading terms
\begin{equation}\label{div}
\frac{d\dot N_\gamma}{dE_\gamma}=\frac{8\pi r^2 c}{\zeta}\int^{1-1/\zeta}_0 \left[1+\frac{x^2}{2(1-x)}\right]F_e(E_{\gamma}/x)dx.
\end{equation}
Here we have approximated the upper limit of the integration by $1-1/\zeta$
because the integral in eq. (\ref{div}) diverges logarithmically at $x=1$.
For an electron distribution with exponential cut-off the main
contribution to the integral comes from regions of $x$ close to unity, so that
$1\ll \frac{x^2}{2(1-x)}$ and we can neglect the first term from the expression
in the brackets,
\begin{equation}\label{corr}
\frac{d\dot N_\gamma}{dE_\gamma}=\frac{4\pi r^2 c}{\zeta}\left[\int^{1-1/\zeta}_0 \frac{x^2 dx}{(1-x)}F_e(\frac{E_\gamma}{x})\right].
\end{equation}
Now the integral of eq. (\ref{corr}) can be calculated for energies
close to the cut-off region, $E_{\gamma}\gg E_c$,
leading to a spectrum of the form
\begin{equation}\label{KNmono}
\left.\frac{d\dot N_\gamma}{dE_\gamma}\right|^{\;KN}=\frac{\pi r^2 c (mc^2)^2}{\varepsilon_{0}E_{\gamma}}\left[\ln\left(\frac{4\varepsilon_{0}E_{\gamma}}{(mc^2)^2}\right)-\ln\beta-\beta\ln(\frac{E_{\gamma}}{E_{c}})-\gamma\right]F_e(E_{\gamma}), \quad {\rm for}  \quad E_{\gamma}\gg E_{c},
\end{equation}
where $\gamma=0.5772$ is the Euler' s constant. In the Klein-Nishina
regime where electrons loose almost all their energy in each scattering,
the Compton Spectrum practically reflects the behavior of the electron
distribution. Thus, in this case, the exponential cut-off maintains the index
$\beta$ and is steeper than in the Thomson case
\begin{equation}\label{KNexp}
\left.\frac{d\dot N_\gamma}{dE_\gamma}\right|^{\;KN}\propto \exp\left[-\left(\frac{E_{\gamma}}{E_{c}}\right)^\beta\right].
\end{equation}
As before, the photon cut-off energy corresponds to the maximum
photon energy that electrons of energy $E_c$ radiate in the Klein-Nishina regime.
Moreover, an abrupt electron distribution cut-off
would result in an abrupt photon spectrum cut-off.
The asymptotics of eq. (\ref{KNmono}) are plotted in figures \ref{f1KN} and \ref{f2KN};
as can be seen, they provide a good approximation just after the peak of the SED.

\begin{figure}[!h]
\epsscale{0.3pt}
\includegraphics[width=250pt,angle=-90]{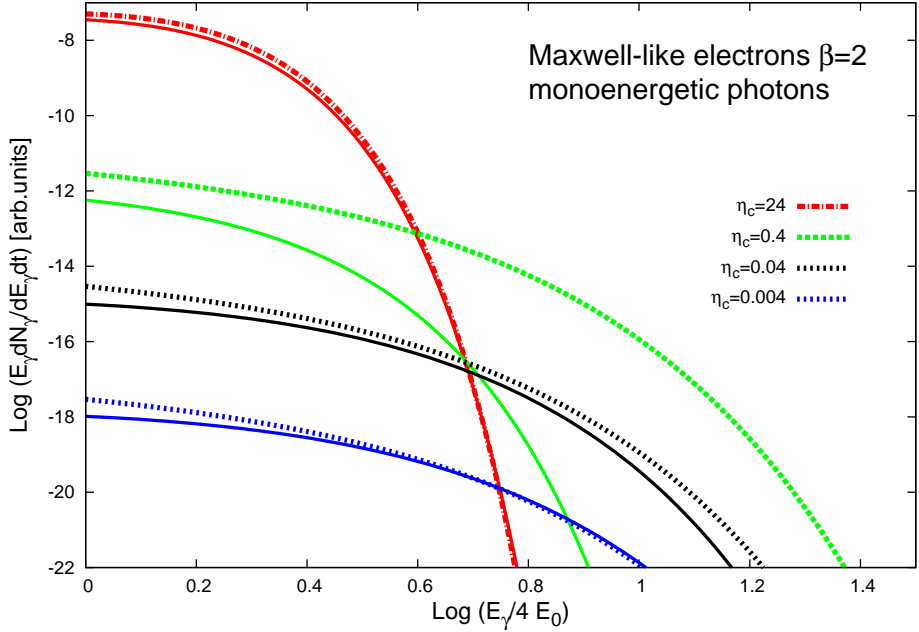}
\caption{Asymptotics for a Maxwell-like distribution with $\beta=2$ for different values of the parameter $\eta_c$ that defines the domain of the scattering for electrons with energy $E_c$. In the Klein-Nishina regime the approximation is good even for not very large values of $\eta_c$. In the Thomson regime the approximation becomes acceptable for $\eta\sim 0.004\ll 1$. In the intermediate domain the exact numerical spectrum decays more sharply than the approximated one, indicating that while $4 \varepsilon_0 E_c<1$, part of the electrons with $E_e>E_c$ already up-scatter the soft photon in the Klein-Nishina regime.}\label{f3}
\end{figure}

The fact that the exponential index becomes $\beta/2$ and $\beta$ in the
Thomson and Klein-Nishina regime, respectively, indicates that using a $\delta$-function
approximation for the cross section provides a correct result for the calculated
spectrum in these two regimes (for an extended discussion on the applicability of
$\delta$-function approximation see \citealt{coppi90}). Obviously this is not true for values of $\eta_c$
close to unity (see fig. \ref{f3}), where $\eta_c=4\varepsilon_{0}E_c$ refers
to the electron cut-off energy. As we are interested in the highest energy part
of the spectrum, the approximation in the Klein-Nishina regime is satisfactory
even for values of $\eta_c$ that do not significantly exceed unity. This happens
because  since $\eta_c>1$ for the electron cut-off energy it holds for all the
energies $E_e>E_c$ that actually form the shape of the exponential cut-off.
On the contrary, for the Thomson regime one needs all the radiated photons above
the cut-off to be emitted at this regime, which indicates rather small values
of $\eta$ for the approximation to be good, especially for low $\beta$ factors.

A similar calculation can be performed for a "pure" power-law
electron distribution, with $\beta=0$.
If we rewrite the second term of eq. (\ref{div}) in the form
\begin{equation}
\int^{1-1/\zeta}_{0}\frac{x^2}{1-x}x^{-a}F_e(E_{\gamma}/x)=F_e(E_{\gamma})\int^{1-1/\zeta}_{0}\frac{x^2 dx}{1-x}+F_e(E_{\gamma})\int^{1}_{0}\frac{(x^{2+\alpha}-1) dx}{1-x},
\end{equation}
we can perform the above integration resulting in an emitted spectrum of the form
\begin{equation}\label{plKN}
\left.\frac{d\dot N_\gamma}{dE_\gamma}\right|^{\;KN}=\frac{\pi r^2 c A(mc^2)^2}{\varepsilon_{0}}E^{-\alpha-1}_{\gamma}\left[\ln\left(\frac{4\varepsilon_{0}E_{\gamma}}{(mc^2)^2}\right)-\frac{3\alpha^2+15\alpha+14}{4\alpha(\alpha+1)(\alpha+2)}-\gamma-\Psi(\alpha)\right],
\end{equation}
where $\Psi(\alpha)$ is the digamma
function defined as the logarithmic derivative of the $\Gamma$
function, $\Psi(\alpha)=\Gamma'(\alpha)/\Gamma(\alpha)$. This formula shows that
the emitted compton spectrum is much steeper in this than in the Thomson regime
due to the suppression of the cross-section. The functional dependance of eq. (\ref{plKN})
\begin{equation}
\left.\frac{d\dot N_\gamma}{dE_\gamma}\right|^{\;KN}\propto E^{-\alpha-1}
\end{equation}
has been obtained in \cite{BG70}, see e.g. their eq. (2.87), and \cite{AA81c}, see their eq. (32). These formulas
differ in the term related to the power-law index $\alpha$, due to the different
approach used for the calculation of the asymptotics. This difference is negligible.

\section{Compton spectrum for a broad photon distribution}

Once we have calculated the radiated spectrum for monochromatic photons,
we can examine the behavior of the Compton spectrum for various photon fields.
The case of monochromatic photons is important for understanding
the scattering mechanism and a necessary step for further calculations.
However, in nature the photon fields are usually broader than the monochromatic one,
except if we deal with emission lines.
Here we will consider a Planckian photon distribution which is often the case in
external Compton scenarios and we will examine as well the case of synchrotron photons
from the same parent electron distribution, that are used as the target
photon field in synchrotron self-Compton models.

\subsection{Planckian photon field}
Let as assume a Planckian distribution for the photon field so that the differential number density is given by
\begin{equation}\label{ph}
n_{ph}(\epsilon_{\gamma})=\frac{1}{\pi^{2}\hbar^{3}c^{3}}\frac{\epsilon^{2}_{\gamma}}{e^{\epsilon_{\gamma}/kT-1}}.
\end{equation}
For the Thomson regime, where now we demand $4kT E_e\ll 1$, we can use the Wien limit ($\epsilon_{\gamma}\gg kT$) at which
\begin{equation}\label{ph}
n_{ph}(\epsilon_{\gamma})=\frac{1}{\pi^{2}\hbar^{3}c^{3}}\epsilon^{2}_{\gamma}e^{-\epsilon_{\gamma}/kT}.
\end{equation}
This is acceptable as
the asymptotic behavior of the Compton spectrum
at high energies is mostly defined by the soft photons with energy
around and greater than $kT$.

In that case the integration of the spectrum eq. (\ref{th3}) over the photon energies
can be evaluated by the saddle point method of integration. After replacing $\varepsilon_0$ with $\epsilon_\gamma$ in eq. (\ref{th3}),
the integral is written as
\begin{equation}
\left.\frac{d\dot N_\gamma}{dE_\gamma}\right|^{T}_{\;BB}=\int^{\infty}_{0}\left.\frac{d\dot N_\gamma}{dE_\gamma}\right|^{T}n_{ph}(\epsilon_{\gamma})d\epsilon_{\gamma}=\int^{\infty}_{0}g(E_{\gamma},\epsilon_{\gamma})e^{S(E_{\gamma},\epsilon_{\gamma})}d\epsilon_{\gamma},
\end{equation}
where
\begin{equation}
g(E_{\gamma},\epsilon_{\gamma})=\frac{8\pi r_e^2 A}{2^{\alpha}\pi^{2}\hbar^{3}c^{2}}\,
\frac{E^{\frac{\alpha-1}{2}}_{\gamma}\epsilon^{\frac{3-\alpha}{2}}_{\gamma}}{\beta^2\xi^2}
\end{equation}
and
\begin{equation}
S(E_{\gamma},\epsilon_{\gamma})=-\left(\frac{E_\gamma}
{4\epsilon_\gamma E_{\rm c}^2}\right)^{\!\beta/2}-\frac{\epsilon_{\gamma}}{kT}.
\end{equation}
Here BB stands for black body. Let as define the parameter $\xi_1$ that is related to the cut-off of the up-scattered photon spectrum
for the case of monochromatic photons, if we replace $\epsilon_{\gamma}$ with $kT$
\begin{equation}
\xi_1\equiv \frac{E_{\gamma}}{4kT E^{2}_{c}}.
\end{equation}
This parameter simply describes the outgoing photon energy normalized
to the maximum energy which $E_{c}$ electrons radiate when they up-scatter photons of energy $kT$.
One can use the saddle point method for the above integration because the integral
at large energies, $\xi_1\gg 1$, is determined by the soft photon
energy interval around the energy $x_0$ which maximizes/minimizes the function
$S(E_{\gamma},\epsilon_{\gamma})$. The saddle point $x_0$ is at
\begin{equation}
x_0=\frac{\beta}{2}\left(\frac{2\xi_1}{\beta}\right)^{\frac{\beta}{\beta+2}}kT.
\end{equation}
Then, the Thomson spectrum for a Planckian photon distribution is calculated to be $d\dot N_\gamma/dE_\gamma=g(E_{\gamma},\epsilon_{*})\;\exp[-S(E_{\gamma},x_0)]\sqrt{2\pi/-S''(E_{\gamma},x_0)}$
where $S''$ the second derivative of $S$ at the saddle point $x_0$.
After rearranging the terms, we retrieve the following expression
\begin{equation}\label{pla}
\left.\frac{d\dot N_\gamma}{dE_\gamma}\right|^{T}_{\;BB}=\frac{4 \pi r_e^2 A (kT)^{\frac{5}{2}} (mc^2)^{\alpha+1}}{2^{\alpha}\pi^{2}\hbar^{3}c^{2}}\sqrt{\frac{\pi}{\beta+2}}x^{-\frac{\alpha}{2}}_{0}E^{\frac{\alpha-1}{2}}_{\gamma}e^{-\frac{\beta+2}{2}\left(\frac{2\xi_1}{\beta}\right)^{\frac{\beta}{\beta+2}}}, \quad {\rm for}  \quad E_{\gamma}\gg \frac{4kT E^{2}_{c}}{(mc^2)^2},
\end{equation}

\begin{figure}[!h]
\epsscale{0.3pt}
\includegraphics[width=190pt,height=290pt,angle=-90]{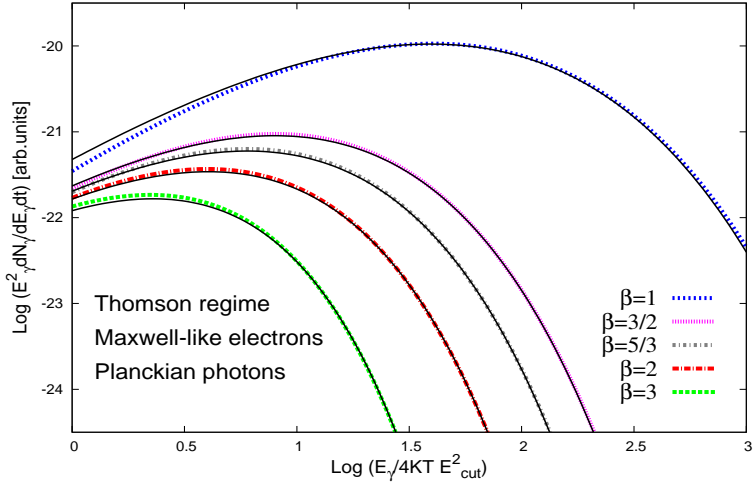}
\caption{Asymptotic behavior of the Thomson spectrum at the cut-off region, for Maxwell-like electrons up-scattering Planckian photons. Parameters used are $E_c=10^{2}$, $KT=10^{-6}$ so that $\eta_c=4 kT E_c=0.0004\ll 1$. The exponential cut-off possesses an index $\beta/(\beta+2)$.}\label{f4T}
\end{figure}

\begin{figure}[!h]
\epsscale{0.3pt}
\includegraphics[width=190pt,height=290pt,angle=-90]{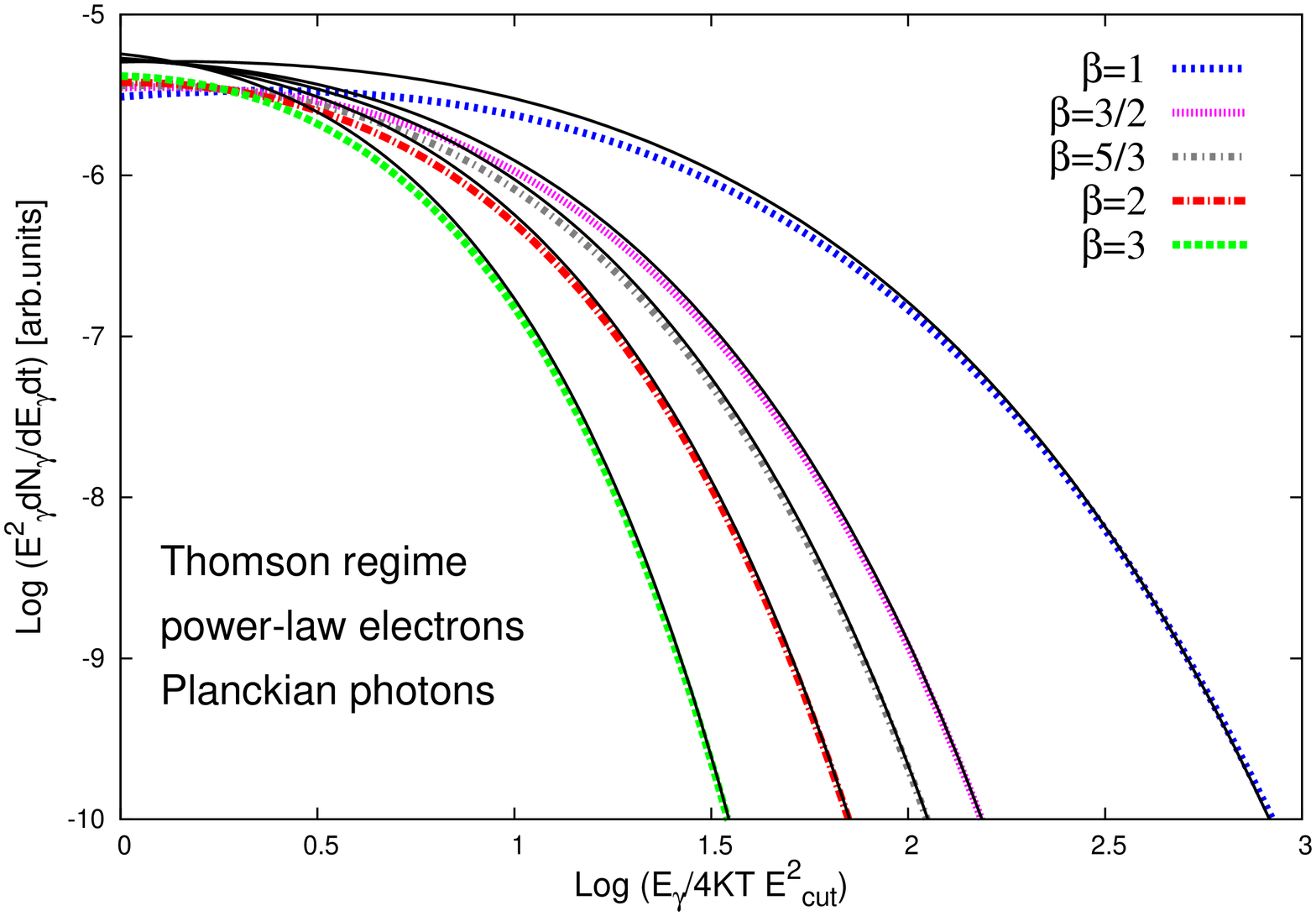}
\caption{Same as figure \ref{f4T}, but for a power-law electron distribution.}\label{f5T}
\end{figure}

Therefore, when the target photon field is a black-body, the shape of
the cut-off is affected by the soft photon distribution and the exponential
cut-off in the Thomson spectrum is smoother in comparison to the monochromatic
photons case. The index now becomes $\beta/(\beta+2)$ as
\begin{equation}\label{BBbeta}
\left.\frac{d\dot N_\gamma}{dE_\gamma}\right|^{T}_{\;BB}\propto \exp{\Big[-\frac{\beta+2}{2}\left(\frac{2}{\beta}\frac{E_{\gamma}(mc^2)^2}{4kT E^{2}_{c}}\right)^{\frac{\beta}{\beta+2}}\Big]}.
\end{equation}
This exponential cut-off is always smooth (less than unity) and it becomes unity in the case of an
abrupt electron distribution cut-off, as $\lim_{\beta\to\infty}{\beta/(\beta+2)}=1$.
Interestingly, the Thomson spectrum for Planckian photons at high energies exhibits the same exponential
cut-off shape as the synchrotron spectrum. This is the only case where the two components of the spectrum show the same
behavior for arbitrary index $\beta$. For the  Maxwellian and
power-law type distributions of electrons, Eq. \ref{BBbeta} is presented
in fig. \ref{f4T} and fig. \ref{f5T} respectively.

In the Klein-Nishina regime the shape of the exponential cut-off does not depend
on the up-scattered photon distribution but, as mentioned above, preserves the electron index $\beta$.
Integration of eq. (\ref{KNmono}) over photon energies (after replacing $\varepsilon_0$ with $\epsilon_\gamma$) requires the calculation of the following integrals
\begin{equation}
\int^{\infty}_0\frac{1}{\epsilon_{\gamma}}n(\epsilon_{\gamma})d\epsilon_{\gamma}=\frac{(kT)^2}{6\hbar c^3},
\end{equation}
\begin{equation}
\int^{\infty}_0\frac{1}{\epsilon_{\gamma}}n(\epsilon_{\gamma})\ln{(4\epsilon_{\gamma} E_{\gamma})}d\epsilon_{\gamma}=\frac{(kT)^2}{6\hbar c^3}\left(\ln{(4kT E_{\gamma})}-0.1472\right).
\end{equation}
Then, the asymptotic behavior of the up-scattered spectrum close to the cut-off follows the formula
\begin{equation}
\left.\frac{d\dot N_\gamma}{dE_\gamma}\right|^{KN}_{\;BB}=\frac{\pi^2 r_e^2 A (KT)^2 (mc^2)^2}{\hbar^3 c^2 }\frac{F_e(E_\gamma)}{E_\gamma}\left[\ln{\frac{4KTE_\gamma}{(mc^2)^2}}-\ln{\beta}-\beta\ln{\frac{E_\gamma}{E_c}}-0.724\right],\quad {\rm for}  \quad E_{\gamma}\gg E_{c}.
\end{equation}

\begin{figure}
\epsscale{0.3pt}
\includegraphics[width=190pt,height=290pt,angle=-90]{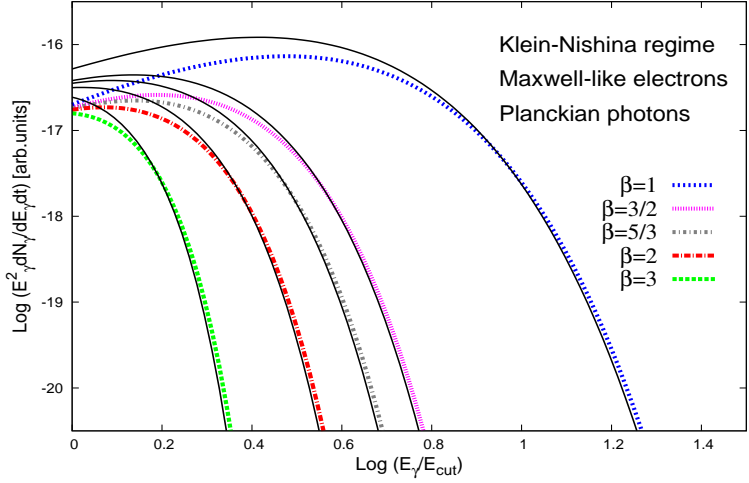}
\caption{Compton spectrum in the Klein-Nishina regime for Maxwellian electron and Planckian photons. $E_c=10^{6}$, $Kk=10^{-4}$ and $\eta_c=4 kT E_c=400\gg 1$. The spectrum shows a cut-off of index $\beta$, same as the electron energy distribution.}\label{f4KN}
\end{figure}

\begin{figure}
\epsscale{0.3pt}
\includegraphics[width=190pt,height=290pt,angle=-90]{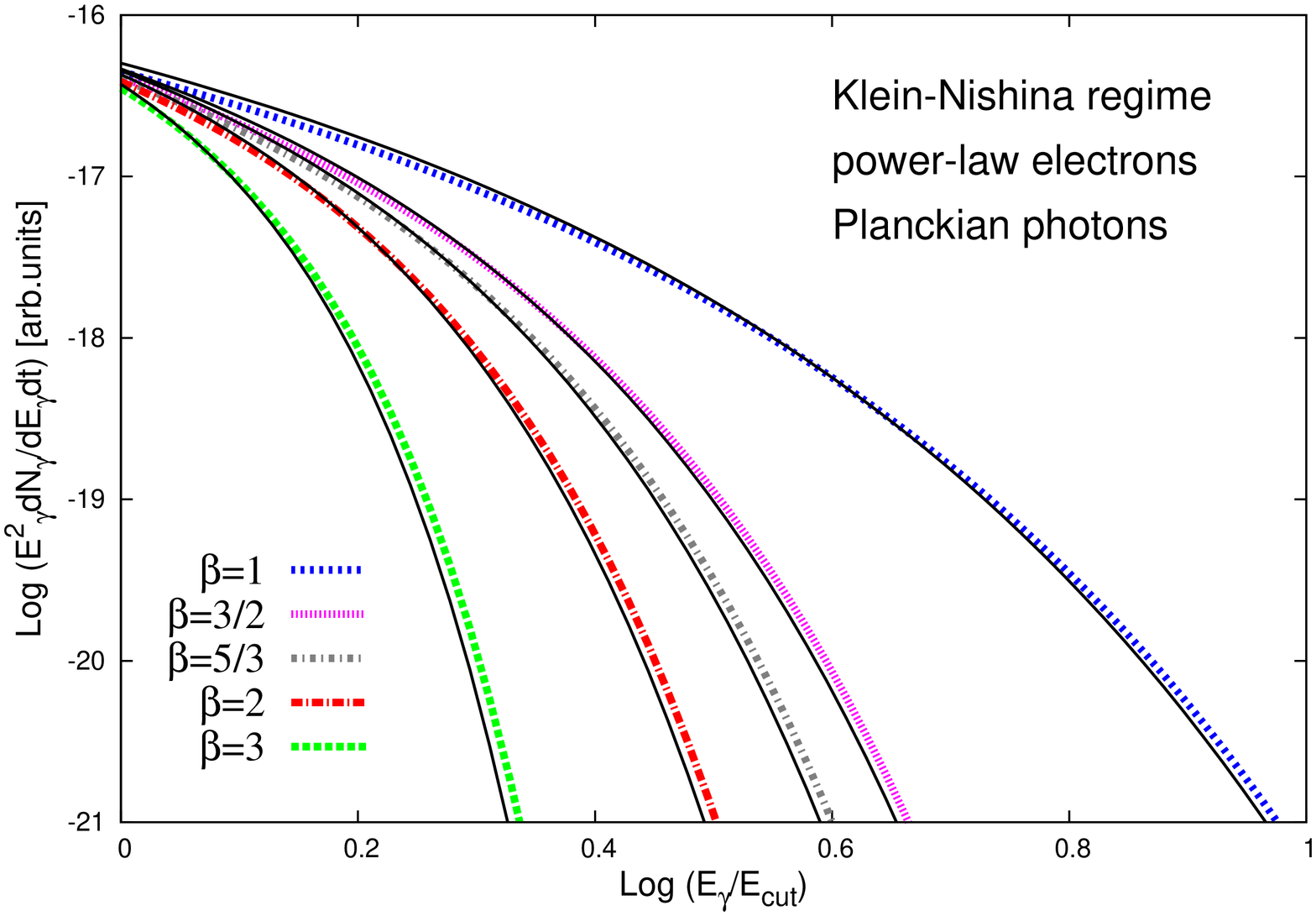}
\caption{Same as fig.\ref{f4KN}, but power-law electrons.}\label{f5KN}
\end{figure}

Thus, in the Klein-Nishina regime the cut-off is always much sharper
than in the Thomson regime. The spectra for Maxwellian and power-law
distributions of electrons are shown in fig. \ref{f4KN} and fig. \ref{f5KN} respectively. For values of the index $\beta=1,2,3$, the corresponding shape in Thomson regime
becomes $1/3$, $1/2$ and $3/5$ respectively, always less than unity. For an abrupt cut-off ($\beta\rightarrow\infty$),
the Klein-Nishina spectrum appears sharp as well, while the Thomson spectrum exhibits a simple exponential cut-off.

\subsection{Synchrotron photon field}
In synchrotron self-Compton models, the electrons up-scatter the photon which they produced via synchrotron radiation.
In contrast to external Compton models, we do not have an analytic expression for the target photon density.
However, we can use an approximation
for the synchrotron spectrum at energies around the synchrotron cut-off,
i.e. for $\epsilon_{\gamma}\geq b E^{2}_{c}$, given that the main contribution
to the scattering process at high energies comes from this energy range. Here $b=3 q B h/4\pi m c (mc^2)^2$.

\subsection{Synchrotron spectrum}
In the case of chaotic magnetic fields, the synchrotron emissivity of an electron with energy $E_e$ is described by the equation
\begin{equation}\label{synformula}
\frac{d\dot{N}_{\gamma}}{d\epsilon_{\gamma}}=\frac{\sqrt{3} q^3 B}{m c^2 h \epsilon_{\gamma}}\tilde{G}\left(\frac{\epsilon_{\gamma}}{\epsilon_{s}}\right),
\end{equation}
where
\begin{equation}\label{Es}
\epsilon_s=b\gamma^2=\frac{3 q B h}{4\pi m c}\frac{E^{2}_{e}}{(mc^2)^2}
\end{equation}
is the "critical" energy for synchrotron emission.
The synchrotron power $\epsilon_{\gamma} \dot{N}_{\gamma}/d\epsilon_{\gamma}$ peaks at $0.29 \epsilon_s$ (e.g. \citealt{RL79})
The function $\tilde{G(y)}$ can be well approximated,
with an accuracy better than 0.2$\%$ over the entire range of variable $y$, by the formula (\citealt{kelner})
\begin{equation}
\tilde{G}(y)=G(y)e^{-y}=\frac{1.808 y^{1/3}}{\sqrt{1+3.4 y^{2/3}}}\frac{1+2.21 y^{2/3}+0.347 y^{4/3}}{1+1.353 y^{2/3}+0.217 y^{4/3}}\;e^{-y}.
\end{equation}
For large $y\gg 1$, the function $G(y)$ is approximately $G(y)\approx\sqrt{\pi/2}$.
Let us define the parameter $\xi_{2}$ that defines the synchrotron photon energy normalized to the energy $\epsilon_s$
(eq. \ref{Es}) for the electrons with energy $E_c$
\begin{equation}
\xi_2=\epsilon_{\gamma}/\epsilon_{0}=\frac{\epsilon_{\gamma}}{b E^{2}_{c}}.
\end{equation}
In order to calculate the emitted synchrotron spectrum, we need to integrate eq. (\ref{synformula}) over the electron distribution given by eq. (\ref{electrons}). By changing variables from $E_e$ to $y=\epsilon_{\gamma}/\epsilon_s$, we find
\begin{equation}
\left.\frac{d\dot{N}_{\gamma}}{d\epsilon_{\gamma}}\right|_{\;SYN}=\int^{\infty}_{0}\frac{d\dot{N}_{\gamma}}{d\epsilon_{\gamma}}F_e(E_e)dE_e=\frac{\sqrt{3} q^3 B A}{2 m c^2 h \epsilon_{\gamma}}(\frac{\epsilon_{\gamma}}{b})^{\frac{\alpha+1}{2}}\int^{\infty}_{0} y^{-\frac{\alpha}{2}-1}G(y)e^{-y-(\frac{\xi_2}{y})^{\beta/2}}dy.
\end{equation}
For $\xi_2>1$ the integration over $y$ reveals a saddle point at
\begin{equation}
y_0=\frac{\beta}{2}\left(\frac{2 \xi_2}{\beta}\right)^{\frac{\beta}{\beta+2}},
\end{equation}
so that finally the emitted synchrotron spectrum can be expressed as
\begin{equation}\label{synchr}
\left.\frac{\epsilon_{\gamma}d\dot{N}_{\gamma}}{d\epsilon_{\gamma}}\right|_{\;SYN}=\frac{\sqrt{3} q^3 B}{m c^2 h}\sqrt{\frac{\pi}{\beta+2}} A \left(\frac{\epsilon_{\gamma}}{b}\right)^{\frac{\alpha+1}{2}}y^{-\frac{\alpha}{2}-1}_0G(y_0) e^{-\frac{2+\beta}{\beta}y_0}.
\end{equation}

\noindent This formula indicates the shape of the exponential cut-off for synchrotron radiation
\begin{equation}\label{SYNbeta}
\left.\frac{\epsilon_{\gamma}d\dot{N}_{\gamma}}{d\epsilon_{\gamma}}\right|_{\;SYN}\propto\exp\left[-\frac{\beta+2}{2}\left(\frac{2\epsilon_{\gamma}}{\beta \epsilon_0}\right)^{\beta/(\beta+2)}\right].
\end{equation}
We note that the $\beta/(\beta+2)$ index for the synchrotron exponential cut-off has already been found
in \cite{frit89} and \cite{zirak07}. With the above calculations we can moreover estimate
critical aspects of the emitted spectrum, and in particular the cut-off energy of the emitted synchrotron
spectrum (see section 4). The above equations correspond to optically thin synchrotron sources when the synchrotron-self absorption can be ignored. This
could be the case of even very compact and highly magnetized sources, as long as the synchrotron cut-off appears at optical and higher frequencies. One should also mention that the synchrotron spectrum is sensitive to inhomogeneities of the magnetic  field (e.g. \citealt{katz94}, \citealt{eilek97}). Obviously, the fluctuations of the magnetic field should have an impact on the synchrotron spectrum, namely they will make the cut-off smoother and shifted towards higher frequencies (see e.g. \citealt{eilek96}). In this regard, the ICS $\gamma$-ray spectrum is free of uncertainties related to the magnetic field distribution, except for realization of the SSC scenario in the Thomson limit.

\begin{figure}
\epsscale{0.3pt}
\includegraphics[width=190pt,height=290pt,angle=-90]{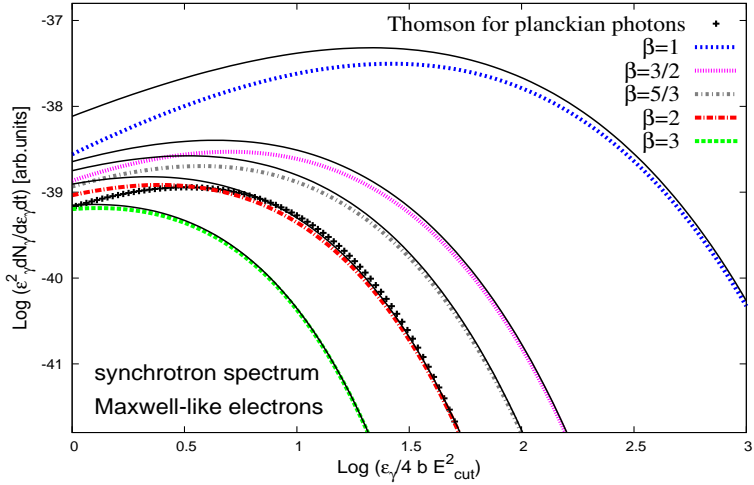}
\caption{Synchrotron spectrum of relativistic Maxwell-like electrons. With black crosses the Thomson spectrum is shown for Planckian photons and $\beta=2$, scaled to $E_{\gamma}/5 kT E^{2}_{cut}$ and normalized to the synchrotron flux.
At large energies, the Thomson and synchrotron spectra show similar shape of the cut-off and the corresponding index is $\beta/(\beta+2)$. The cut-off energy for the electrons is $E_c=10^2$ and the magnetic field $B=1G$.}\label{fsmax}
\end{figure}

\begin{figure}
\epsscale{0.3pt}
\includegraphics[width=190pt,height=290pt,angle=-90]{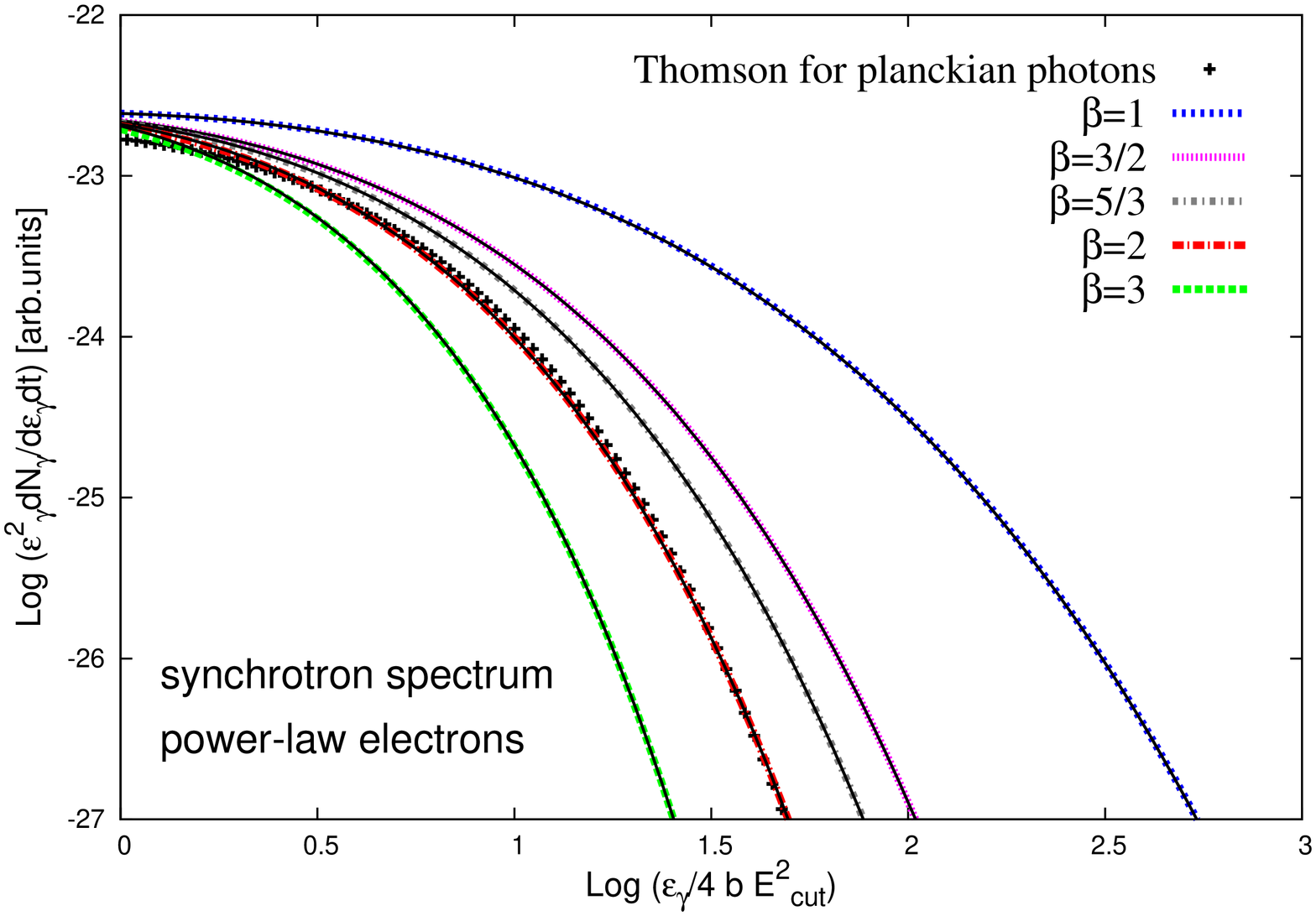}
\caption{Same as fig.\ref{fsmax}, but power-law electrons.}\label{fspl}
\end{figure}

Last, we should mention that
using a $\delta$-function approximation for the synchrotron emissivity
would result in an exponential cut-off of index $\beta/2$, whereas the correct value is $\beta/(2+\beta)$.
The synchrotron asymptotics are plotted in figures \ref{fsmax} and \ref{fspl}, for Maxwellian and power-law
electron distributions, respectively. For comparison, the Thomson spectrum for Planckian photons is shown as well.

\subsection{SSC spectrum}
Now we can integrate the Compton spectrum for monochromatic photons over the photon energies
for the synchrotron distribution. The differential photon number density (for a spherical source) is
\begin{equation}\label{numdens}
n_{ph}(\epsilon_{\gamma})=\frac{3}{4\pi R^3}\frac{R}{c}\frac{d\dot{N}_{\gamma}}{d\epsilon_{\gamma}},
\end{equation}
where R is the source size. Let us first perform the calculations for the Thomson regime described by eq. (\ref{th3}) with $\varepsilon_0\rightarrow\epsilon_\gamma$.
In that case the exponential factor in the integrant is $\exp{[-S]}$, where $S=\xi+(2-\beta)y_0/\beta$.
The function $S$ has extremum at the saddle point
\begin{equation}
z_0=\frac{\beta}{2}\left(\frac{2}{\beta}\frac{E_{\gamma}}{E_0}\right)^{\frac{\beta+2}{\beta+4}}\epsilon_0,\;\;\;\;
\end{equation}
where $E_0=4 b E^{4}_{c}$ is the maximum photon energy that results from electrons with energy $E_c$ when they up-scatter synchrotron photons with energy $\epsilon_0$. The analogy with the case of Planckian soft photons is direct. Replacing the synchrotron characteristic energy $b E^2_{c}$ with $kT$ results in the same saddle point.

Here the second derivative of the exponential argument at the saddle point has a simple form
\begin{equation}
S''(z_0)=\frac{\beta E_0}{2 E_{\gamma}}\frac{1}{\epsilon^{2}_{0}},
\end{equation}
so that the integration over the synchrotron number density gives
\begin{equation}\label{SSCtho}
\left.\frac{d\dot{N}_{\gamma}}{dE_{\gamma}}\right|^{\;T}_{\;SSC}=\frac{3}{4\pi c R^2}\frac{2\pi^2 r^2 c A^2 (mc^2)^{\alpha+1} \sqrt{3}q^3 B}{2^\alpha\sqrt{\beta(\beta+2)} h}\left(\frac{E_{\gamma}}{b}\right)^{\frac{\alpha}{2}}\frac{G(\tilde{y_0})}{z_0} \tilde{y_0}^{-\frac{\alpha}{2}-3}\;e^{-\frac{\beta+4}{\beta}\tilde{y_0}},
\end{equation}
where $\tilde{y_0}$ is calculated at the saddle point
\begin{equation}
\tilde{y_0}=\frac{\beta}{2}\left[\frac{2 E_{\gamma}}{\beta E_0}\right]^{\frac{\beta}{\beta+4}},\;\;\;\;E_0=\frac{4 b E^{4}_{c}}{(mc^2)^2}.
\end{equation}

\begin{figure}
\epsscale{0.3pt}
\includegraphics[width=190pt,height=290pt,angle=-90]{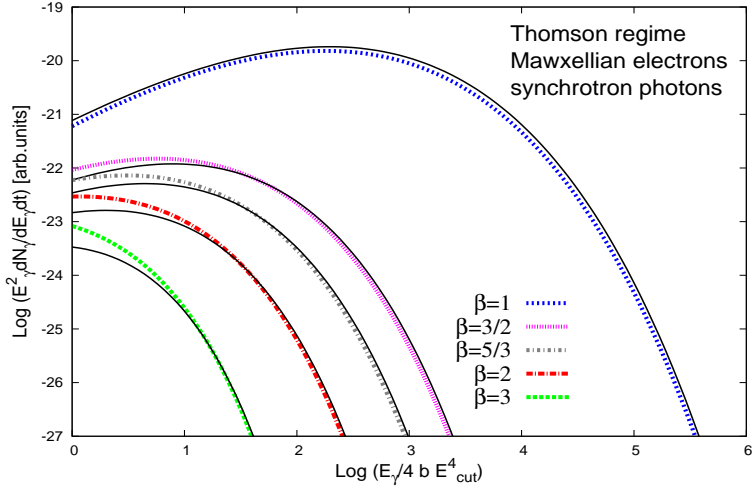}
\caption{SSC radiation at the cut-off region in Thomson regime. A Maxwellian electron distribution has been used, with parameters $E_c=10^2$ and $B=1G$ for the magnetic field. The exponential cut-off that arises is very smooth, of index $\beta/(\beta+4)$}\label{f7T}
\end{figure}

\begin{figure}
\epsscale{0.3pt}
\includegraphics[width=190pt,height=290pt,angle=-90]{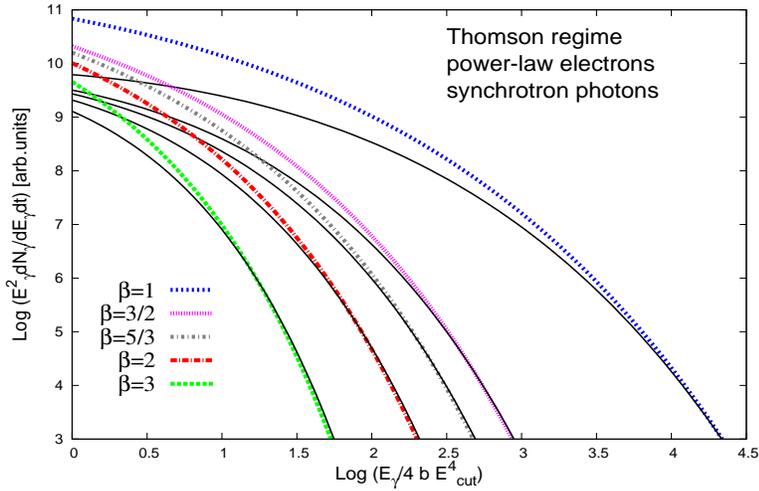}
\caption{Same as fig.\ref{f7T} but power-law electrons. The asymptotics approach the numerical solution only for $E_{\gamma}\gg 4b E^4_{cut}$. Very close to the photon cut-off energy, the numerical spectrum is smoother than the approximated one.}\label{f6T}
\end{figure}

In the case of SSC radiation, the electron distribution up-scatters the synchrotron photon distribution with an exponential cut-off $\exp[-(\epsilon_{\gamma}/\epsilon_{c})^{\frac{\beta}{\beta+2}}]$. The corresponding Compton flux at high energies exhibits a cut-off index $\beta/(\beta+4)$ which is smoother than the seed synchrotron distribution,
\begin{equation}\label{SSCbeta}
\left.\frac{E_{\gamma}d\dot N_\gamma}{dE_\gamma}\right|^{\;T}_{\;SSC}\propto \exp\Big[-\frac{\beta+4}{2}\left(\frac{2 E_{\gamma}}{\beta E_0}\right)^{\frac{\beta}{\beta+4}}\Big].
\end{equation}
For $\beta=1,2,3$, the corresponding values for the Thomson exponential index is $1/5, 1/3, 3/7$, significantly less than the electron distribution index and different than the synchrotron case. If $\beta\rightarrow\infty$, then the SSC spectrum shows a simple cut-off,
like in the case of up-scattering Planckian photons. The asympotic formula of eq. (\ref{SSCtho}) is plotted in figures \ref{f7T} and \ref{f6T} for Maxwellian and power-law electrons.

\begin{figure}
\epsscale{0.3pt}
\includegraphics[width=190pt,height=290pt,angle=-90]{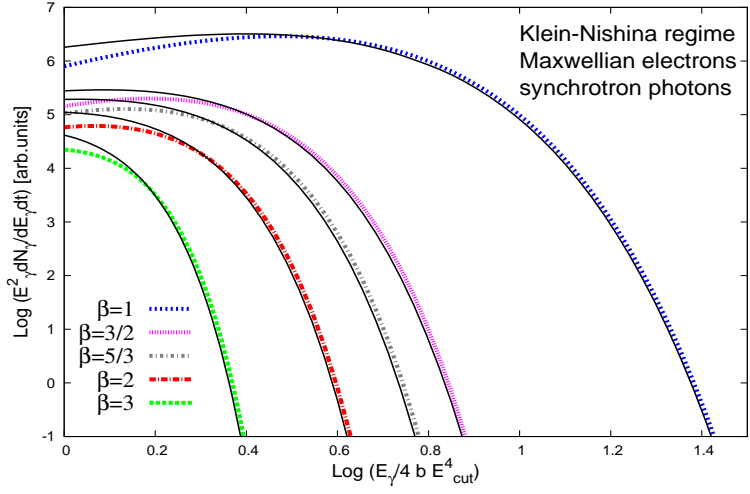}
\caption{SSC radiation at the cut-off region in the Klein-Nishina regime for Maxwellian electrons. The parameters used are $E_{c}=10^5$ and $B=1G$, so that the parameter $\eta_c\approx 4bE^{4}_c$ that defines the domain of the scattering is $\eta_c\approx 90\gg 1$. The analytical formula plotted here is eq. (\ref{th3}) for monoenergetic photons with $\epsilon_{\gamma}=bE^{2}_{cut}$, scaled to the numerical spectrum. The cut-off shape has the same index $\beta$, as the electron distribution.}\label{f7KN}
\end{figure}

\begin{figure}
\epsscale{0.3pt}
\includegraphics[width=190pt,height=290pt,angle=-90]{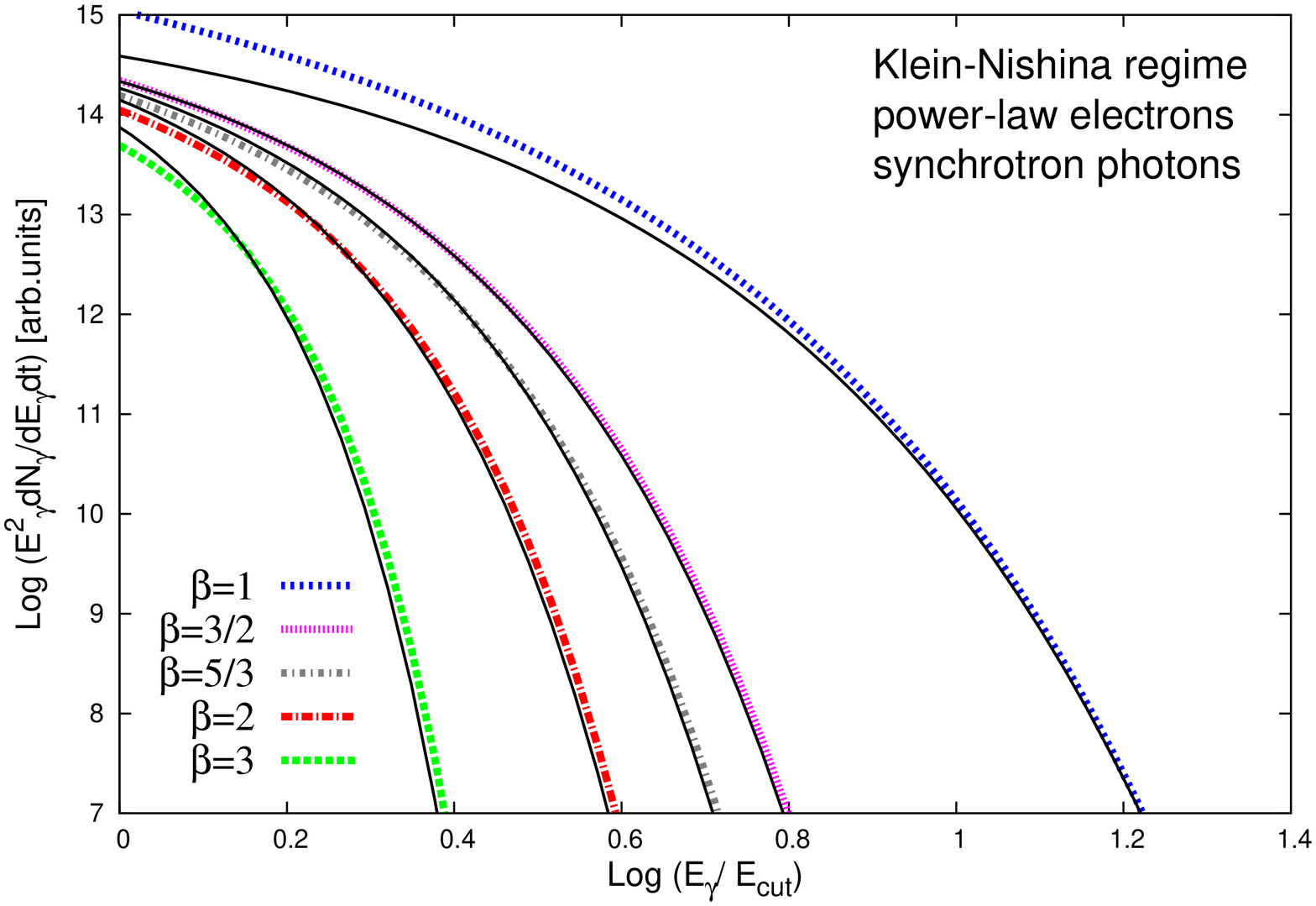}
\caption{Same as fig.\ref{f7KN}, but power-law electrons.}\label{f6KN}
\end{figure}

In the Klein-Nishina regime, the integration over the synchrotron photon density can not be performed analytically for arbitrary values of the indexes $\alpha$ and $\beta$. In this regime however, the soft photon field does not play an important role in the shape of the up-scattered spectrum close to the maximum cut-off. Thus, eq. (\ref{KNmono}) for monochromatic photons offers a rather good description of the asymptotic behavior of the SSC spectrum at Klein-Nishina regime. See e.g. the figures \ref{f7KN} and \ref{f6KN} for Maxwellian and power-law electrons respectively. The analytic formula describing the asymptotes is eq. (\ref{KNmono}) normalized to the numerical solution, where the soft photon energy $\epsilon_{\gamma}$ has been replaced by $\epsilon_s=b E^2_c$.

\section{Discussion}
The main results of this paper are summarized in table~\ref{table1}. In the Klein-Nishina regime the up-scattered Compton spectrum exhibits the same exponential cut-off index $\beta_C$, as the electron distribution index $\beta$, and does not depend strongly on the target photon field. This implies that from the $\gamma$-ray spectrum we practically "observe" the electron cut-off shape. In particular, an abrupt electron distribution cut-off ($\beta\rightarrow\infty$) would result in an abrupt cut-off for the photon spectrum ($\beta\rightarrow\infty$). The case of up-scattering monochromatic photons is shown in figures \ref{f1KN} and \ref{f2KN}. The case of Planckian photons is plotted in figures \ref{f4KN} and \ref{f5KN}, whereas figures \ref{f7KN} and \ref{f6KN} correspond to the SSC spectrum. In all of these cases, the asymptotics are rather good just after the peak of the SED.

On the contrary, in the Thomson regime the up-scattered photon exponential shape is always smoother than the electron distribution cut-off shape. For monochromatic target photons this is $\beta_C=\beta/2$, as shown in figures \ref{f1T} and \ref{f2T}. In this case, the shape of the cut-off in both Thomson and Klein-Nishina regime (eqs. \ref{beta2} and \ref{KNexp}) shows that using a $\delta$-function for the Compton emissivity provides the correct result. Here the asymptotic analytic expression of eq. (\ref{th3}) approaches better the numerical solution for a Maxwellian electron distribution.

Interestingly, for Planckian photons we find a different relation between $\beta_C$ and $\beta$, which is the same as for synchrotron radiation, $\beta_C=\beta/(\beta+2)$. As can be seen from figures \ref{f4T} and \ref{f5T}, the approximation is very good.  Finally, for synchrotron photons it holds that $\beta_C=\beta/(\beta+4)$. As in the case of monochromatic photons, the SSC asymptotics in the Thomson regime are better for Maxwellian electrons.

In general, although in the Klein-Nishina regime the Compton spectrum preserves the electron distribution index $\beta_C=\beta$, in the Thomson regime the up-scatter photon cut-off index is always smaller than the electron distribution cut-off, $\beta_C< \beta$. The only exception is for monochromatic photons up-scattered by electrons with $\beta\rightarrow\infty$. In this case the Compton spectrum should exhibit as well an abrupt cut-off. For Planckian and synchrotron photons $\beta\rightarrow\infty$ for electrons means a simple exponential cut-off for the Compton SED, $\beta_C=1$.

The Thomson spectrum for Planckian photons exhibits a cut-off index $\beta_S=\beta/(\beta+2)$, same as the synchrotron spectrum. Thus,
if the SED of the observed object is considered to consists of these two components, synchrotron radiation for the low energies and up-scattering of Planckian photons in the Thomson regime for high energies, then the exponential part of the two "bumps" is very similar (see fig.~\ref{fsmax} and \ref{fspl}). If, on the other hand, the up-scattering occurs in Klein-Nishina regime in the energy band of the cut-off, then this is not true. While $\beta_S=\beta/(\beta+2)$ for the synchrotron component cut-off, $\beta_C=\beta$ for the high energy component. Even if we consider an abrupt cut-off for the electron distribution, the two bumps would be different ($\beta_S=1$ and $\beta_C\rightarrow\infty$, respectively). For an SSC model, the two components do not show the same exponential cut-off shape neither in the Thomson (where we get $\beta_S=\beta/(\beta+2)$ and $\beta_C=\beta/(\beta+4)$ for low and high energies respectively), nor in the Klein-Nishina regime ($\beta_S=\beta/(\beta+2)$ and $\beta_C=\beta$). Only if the electron distribution has an abrupt cut-off, then in the Thomson regime we can get $\beta_C=\beta_S=1$, while in the Klein-Nishina regime $\beta_C\rightarrow\infty$. Thus, the two components of the SED do not show in general the same shape at the cut-off region.

Apart from the shape of the up-scattered, photon spectrum at the cut-off region, another interesting point concerns the cut-off energy itself, $E_{\gamma,cut}$. We have shown that in the Klein-Nishina regime this remains always at
$E_{\gamma,cut}=E_{c}$, same as the electron distribution cut-off energy, independently
of the soft photon field.
On the contrary, at the Thomson regime, the cut-off photon energy depends on the target photon field.
 For the monochromatic target photons the value of E... is obvious; it is simply equal to the cutoff energy of electrons.
For monochromatic photons,
the resulting value of $E_{\gamma,cut}$ is rather obvious, simply because there is a maximum up-scattered photon energy
for the fixed electron energy. Consequently, the cut-off energy of the IC spectrum must be equal to the maximum up-scattered photon energy
by electrons of energy $E_c$, as it follows from (eq.~\ref{beta2}),
\begin{equation}
E_{\gamma,cut}=\frac{4E^{2}_{c}\varepsilon_{0}}{(mc^2)^2}.
\end{equation}
On the other hand, in the case of a broad distribution of target photons, the relation between $E_c$ and
$E_{\gamma,cut}$ depends on the seed photon spectrum and the index $\beta$ of the electron distribution.
For the Planckian photon distribution, we find (see eq.~\ref{BBbeta})
\begin{equation}\label{BBfactor}
E^{BB}_{\gamma,cut}=\frac{4E^{2}_{c}kT}{(mc^2)^2}\left(\frac{2}{\beta+2}\right)^{\frac{\beta+2}{\beta}}\frac{\beta}{2}.
\end{equation}
In analogy to the monoenergetic photons case, the maximum energy at which electrons of energy $E_c$ can radiate, when they up-scatter photons of energy $kT$, is $4 E^{2}_{c}KT/(mc^2)^2$. In respect to this, the Thomson spectrum cut-off energy is smaller by a factor of $(\beta/2)[2/(\beta+2)]^{(\beta+2)/2}$. This factor is not negligible especially for small $\beta$,  e.g. it takes values of $\sim 0.15$, $0.25$, $0.33$ for $\beta=1,2,3$ respectively (almost one order of magnitude for a simple exponential cut-off).
As expected, it tends to unity for $\beta\rightarrow\infty$. It does not depend however on index $\alpha$ of the electron distribution.

For synchrotron radiation, from eq.~(\ref{SYNbeta}), we find the cut-off energy
\begin{equation}
\epsilon^{SYN}_{\gamma,cut}=b E^{2}_{c}\left(\frac{2}{\beta+2}\right)^{\frac{\beta+2}{\beta}}\frac{\beta}{2},
\end{equation}
which reveals exactly the same factor as in the case of Thomson spectrum for Planckian photons, but now in respect to the characteristic energy $b E^{2}_{c}$.

Finally, for the case of SSC, eq.~(\ref{SSCbeta}) gives
\begin{equation}
E^{SSC}_{\gamma,cut}=\frac{4bE^{4}_{c}}{(mc^2)^2}\left(\frac{2}{\beta+4}\right)^{\frac{\beta+4}{\beta}}\frac{\beta}{2}=\frac{4E^{2}_{c}\;\epsilon^{SYN}_{\gamma,cut}}{(mc^2)^2}\left(\frac{2}{\beta+4}\right)^{\frac{\beta+4}{\beta}}\left(\frac{\beta+2}{2}\right)^{\frac{\beta+2}{\beta}}
\end{equation}
If we compare the cut-off energy with e.g. $4bE^{2}_{cut}\;\epsilon^{SYN}_{\gamma,cut}/(mc^2)^2$ then the factor related to the index $\beta$ takes values of $\sim 0.035,0.15,0.25$ for $\beta=1,2,3$, slightly less than in the case of a Planckian target photon field. This is due to the fact that the cut-off shape is now smoother. These analytic results are useful when modeling the observed spectrum, as one may infer the electron distribution cut-off energy from the photon spectrum cut-off energy with only the uncertainty introduced by a (possible) Doppler boosting. The position and the amplitude of the synchrotron and IC peaks in the SED contain very important information about physical parameters of non thermal sources, like the strength of the average magnetic field and the
energy density of relativistic electrons. The shape  of the SED, especially in the region of the cutoffs of the synchrotron and IC components of radiation, provide additional, more detailed information about the distributions of electrons and magnetic fields. For example, the spectral cutoff in the IC component formed in the
the Klein-Nishina regime provides direct, model-independent information about the
energy spectrum of highest energy electrons. This is a critical issue for understanding of
particle acceleration mechanisms. Furthermore, combined with the shape of the synchrotron cut-off, it can allow us to extract information about the distribution of the magnetic field.
This can be demonstrated by the following simple example. Let as assume that we have observed a smooth synchrotron cut-off  which can be interpreted as
the result of an electron distribution with an exponential index e.g. $\beta \approx 1$.
This hypothesis can be checked by the shape of the cut-off of the IC component.
If the latter is formed in the Klein-Nishina regime, and exhibits a sharp cut-off behavior indicating
to the electron distribution with  $\beta > 1$, then one should
attribute the smoothness of the synchrotron cut-off to magnetic field inhomogeneities
rather than to the actual shape of the electron distribution.

\begin{table}[!h]
\begin{center}
\centering
\caption{exponential cut-off index for Compton spectrum}\label{table1}
\begin{tabular}{|c|c|c|c|c|c|}
\hline
electron index & $\;\;\;$$\beta$$\;\;\;$ & $\beta$ & $\beta\rightarrow\infty$ & $\beta\rightarrow\infty$\\ \hline
scattering regime & $\;\;\;$Thomson$\;\;\;$ & Klein-Nishina & $\;\;\;$Thomson$\;\;\;$ & Klein-Nishina\\ \hline
monochromatic photons & $\beta/2$ & $\beta$ & $\beta\rightarrow\infty$ & $\beta\rightarrow\infty$\\ \hline
Planckian photons & $\beta/(\beta+2)$ & $\beta$ & 1 & $\beta\rightarrow\infty$\\ \hline
synchrotron photons & $\beta/(\beta+4)$ & $\beta$ & 1 & $\beta\rightarrow\infty$\\ \hline
\end{tabular}
\end{center}
\end{table}

\section{Summary}
In this paper we have examined the asymptotic behavior of the Compton spectrum close to the maximum cut-off.
We assumed that the electron distribution follows the general formula $E^{\alpha}_e \exp[-(E_e/E_{c})^{\beta}]$
so that our analysis may account for a relativistic Maxwellian-type distribution, as well as for a power-law distribution
with exponential cut-off. The exponential cut-off of the electron energy spectrum results in an exponential cut-off in the Compton spectrum, of the form $\exp[-(E_{\gamma}/E_{\gamma,cut})^{\beta_c}]$, with $\beta_c$ and $E_{\gamma,cut}$ the corresponding cut-off index and energy respectively.
We show that in the Klein-Nishina regime, the cut-off index remains unchanged, $\beta_c=\beta$.
The shape of the up-scattered spectrum close to the maximum cut-off basically "reflects" the electron distribution
and does not depend strongly on the target photon field. The cut-off energy also correspond to the electron distribution cut-off energy, $E_{\gamma,cut}=E_c$.

In the Thomson regime, the resulting spectrum close to the cut-off is very different. First of all it strongly depends on the up-scattered photon field. Monoenergetic photons lead to a cut-off of index $\beta_c=\beta/2$, whereas Planckian photons result in $\beta_c=\beta/(\beta+2)$. When the up-scattered photon field is the synchrotron photon field, as in SSC models, then the cut-off appears extremely smooth, with an index $\beta_c=\beta/(\beta+4)$. In contrast to the Klein-Nishina regime, the Thomson spectrum cut-off energy $E_{\gamma,cut}$ depends not only on the electron distribution cut-off energy, but also on the target photon field and as well on the index $\beta$.

The obtained analytic expressions are useful for deriving the electron spectral shape at the cut-off region directly from the observed high energy flux. These two parameters may give important insight into the acceleration and radiation mechanisms acting in the source. Furthermore, one may use the higher energy part of the observed Compton spectrum as a "diagnostic tool" to distinguish between EC and SSC models, as different photon fields lead to different cut-off shapes.

{\it Acknowledgement: We would like to thank F. M. Rieger for helpful discussions and careful reading of the manuscript.}

\clearpage

{}

\end{document}